# Optical Absorption and Electron Paramagnetic Resonance of the E'$_\alpha$ Center in Amorphous Silicon Dioxide

G. Buscarino,* R. Boscaino, S. Agnello, and F. M. Gelardi

*Department of Physical and Astronomical Sciences, University of Palermo, Via Archirafi 36, I-90123 Palermo, Italy*

We report a combined study by optical absorption (OA) and electron paramagnetic resonance (EPR) spectroscopy on the E'$_\alpha$ point defect in amorphous silicon dioxide (a-SiO$_2$). This defect has been studied in β-ray irradiated and thermally treated oxygen-deficient a-SiO$_2$ materials. Our results have pointed out that the E'$_\alpha$ center is responsible for an OA Gaussian band peaked at ~ 5.8 eV and having a full width at half maximum (FWHM) of ~ 0.6 eV. The estimated oscillator strength of the related electronic transition is ~ 0.14. Furthermore, we have found that this OA band is quite similar to that of the E'$_\gamma$ center induced in the same materials, indicating that the related electronic transitions involve states highly localized on a structure common to both defects: the O≡Si$^\bullet$ moiety.

## I. INTRODUCTION

Amorphous silicon dioxide (a-SiO$_2$) is a material largely used in many modern and relevant applications [1,2]. Its high transparency in the visible-UV range, for example, motivates its large use to produce a wide variety of optical devices [1,2]. The drawback of using this material is connected with the fact that the performances of the devices are largely influenced by the presence of point defects, induced by exposition to ionizing radiation, which bring electronic states within the gap of forbidden energies of a-SiO$_2$, reducing the transparency of the material [1,2].

In this context a key role is played by the E'$_\gamma$ center [1,2], whose most accepted microscopic model is schematically represented in Fig. 1(a) and consists in a positively charged oxygen vacancy: O≡Si$^\bullet$ $^+$Si≡O (where ≡ represents the bonds to three oxygen atoms, $^\bullet$ is an unpaired electron in a Si-$sp^3$ orbital and $^+$ is a trapped hole) [1-5]. The attribution of this microscopic model to the defect is mainly due to the characterization of its $^{29}$Si hyperfine structure by electron paramagnetic resonance (EPR) spectroscopy [3]. The E'$_\gamma$ center influences the optical properties of a-SiO$_2$ through an optical absorption (OA) Gaussian band peaked at 5.75 ÷ 5.85 eV and having a full width at half maximum (FWHM) of 0.6 ÷ 0.8 eV [1,6-9]. This variability depends on the irradiation dose and on the thermal treatments which the material has been subjected to [9]. Although the attribution of this OA band to the E'$_\gamma$ center is at present considered certain, the degree of localization of the states involved in the related electronic transition is a matter of debate since 1980 [1,8]. In particular, two distinct schemes have been proposed [1,8]. In the former [10-12] both the ground and the first excited electronic states are highly localized on the O≡Si$^\bullet$ moiety, whereas in the latter [13-15] the transition involves the transfer of the unpaired electron from the Si-$sp^3$ orbital of the O≡Si$^\bullet$ molecular group to a Si orbital of the opposite structure, $^+$Si≡O.

Another relevant point defect falling into the class of the so called E' family [1,2,4,16-19] in a-SiO$_2$ is the E'$_\alpha$ [1,4,18-20]. In contrast with the E'$_\gamma$ center, very little was known about it for a long time. Only recently it has been proven that its $^{29}$Si hyperfine structure comprises a pair of EPR lines split by ~49 mT [18,19], suggesting that the E'$_\alpha$ center consists in a positively charged oxygen vacancy with the unpaired electron Si-$sp^3$ orbital pointing away from the vacancy in a back-projected configuration and interacting with an extra O atom of the a-SiO$_2$ matrix, as shown in Fig. 1(b). On the basis of this microscopic model, it is inferred that the unpaired electrons involved in E'$_\gamma$ and E'$_\alpha$ centers are located on quite similar O≡Si$^\bullet$ molecular groups [compare Fig. 1(a) and 1(b)]. Although those works [18,19] have provided fundamental information on the atomic scale structure of the E'$_\alpha$ center, other relevant questions concerning this defect remain open. In particular, in spite of the key role played by a-SiO$_2$ in a wide variety of modern applications, the influence of the E'$_\alpha$ center on the optical properties of this technologically relevant material has never been clarified.

In order to obtain insight into this topic we have performed a combined study by OA and EPR spectroscopy on the E'$_\alpha$ center in oxygen-deficient a-SiO$_2$ materials subjected to β-ray irradiation and thermal treatments. The results we report allow to identify and characterize the OA band of the E'$_\alpha$ center for the first time. Furthermore, our data point out that this OA band is quite similar to that of the E'$_\gamma$ center, indicating that the related electronic transitions involve states highly localized on the same O≡Si• moiety.

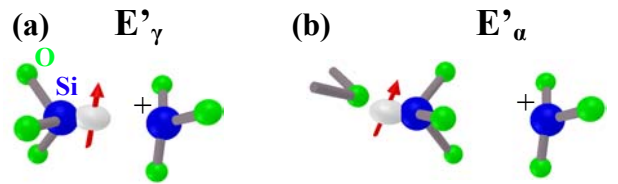

FIG. 1.  (color online). Microscopic structures proposed for (a) E'$_\gamma$ [1-5] and (b) E'$_\alpha$ [18,19] centers in a-SiO$_2$. Arrows represent unpaired electrons in Si-$sp^3$ orbitals and + indicates a trapped hole.

## II. EXPERIMENTAL DETAILS

The materials considered here are commercial a-SiO$_2$. Two of them are obtained from fused quartz, QC and Puropsil A (QPA) [21], while a third material, KUVI [22], is synthesized via vapour axial deposition. Samples are rectangular shaped with 5 mm x 5 mm optically polished surfaces and thicknesses: 0.6 mm (QC and QPA) and 1 mm (KUVI). One sample of each material was preliminarily β-ray irradiated in a Van de Graff accelerator at room temperature at a dose of ~5x10$^3$ kGy. By performing EPR measurements on the irradiated samples an initial concentration of E'$_\gamma$ centers of 1x10$^{18}$ spins/cm$^3$ in QPA, 1.4x10$^{17}$ spins/cm$^3$ in QC and 4.8x10$^{16}$ spins/cm$^3$ in KUVI was estimated, while no E'$_\alpha$ centers were detected. Furthermore, with the same technique a concentration of [AlO$_4$]$^0$ centers of ~3x10$^{17}$ spins/cm$^3$ was estimated in QC and KUVI, whereas it was below the detection limit in QPA.

Irradiated samples were thermally treated at T=630 K for ~20 hours. As described in details elsewhere [18,19], in Al containing a-SiO$_2$ materials, as QC and KUVI, this treatment activates hole transfer processes from the [AlO$_4$]$^0$ centers to the precursors sites of the E' centers, generating E'$_\gamma$ and E'$_\alpha$ centers. At variance, in the QPA sample, in which [AlO$_4$]$^0$ centers were not induced by irradiation, the same thermal treatment is unable to generate E' defects and only a partial annealing of the already present E'$_\gamma$ centers is observed.

Finally, to study the contribution of the E'$_\alpha$ centers to the OA properties of a-SiO$_2$, the three samples were subjected to 25 min isochronal thermal treatments at increasing temperature from T=640 K to T=860 K by steps of 10 K. As a consequence of these treatments, the concentrations of E' centers gradually decrease in all the considered samples. Our experiment consisted in looking for correlations between the changes induced by these treatments in the concentrations of E'$_\alpha$ and E'$_\gamma$ centers and those occurring in the OA spectrum. In particular, the concentrations of E'$_\alpha$ and E'$_\gamma$ centers were estimated after each thermal treatment by EPR measurements applying the decomposition procedure described in Refs. 18 and 19, whereas OA spectra were acquired only after the thermal treatments at T= 630 K, T=710 K, T=740 K, T=780 K and T=820 K.

OA spectra were acquired in the range 4 ÷ 8.2 eV at room temperature with an ACTON vacuum-UV spectrometer (Mod. SP150) working in N$_2$ flux (typically 80 l/min). Experimental spectra were corrected for the reflection from sample surfaces by using literature data on the refractive index dispersion in a-SiO$_2$ [23]. EPR measurements were carried out at room temperature with a Bruker EMX spectrometer working at frequency ν ≈ 9.8 GHz (X-band), with magnetic-field modulation frequency of 100 kHz and acquiring in the first-harmonic unsaturated mode. The concentration of defects was determined by comparing the double integral of the EPR spectrum with that of a strong pitch standard (0.11% pitch in KCl) from Bruker, taking into account differences in filling factors. The estimated accuracy of the absolute concentration is ±50%, whereas that of the relative concentration is ±10%.

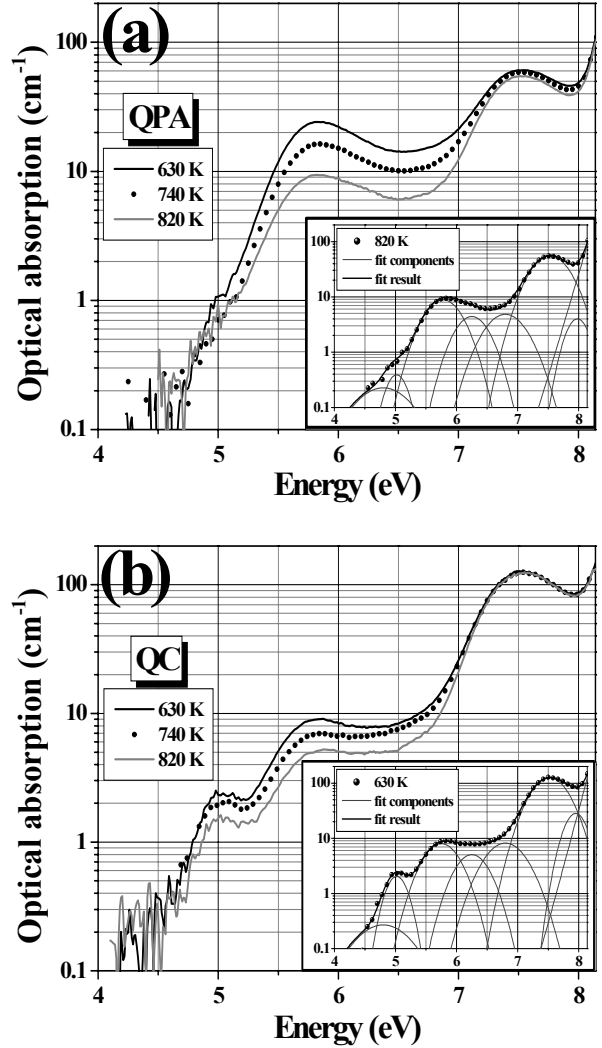

FIG. 2. Comparison among the experimental OA spectra obtained after irradiation and thermal treatment at T=630 K, T=740 K and 820 K for the samples (a) QPA and (b) QC. Insets: Fit of the OA spectra obtained for the samples (a) QPA after irradiation and thermal treatment at T=820 K and (b) QC after irradiation and thermal treatment at T=630 K.

## III. EXPERIMENTAL RESULTS

The OA spectra obtained for T=630 K, T=740 K and T=820 K are compared in Fig. 2(a) and 2(b) for the samples QPA and QC, respectively. As shown, on increasing the thermal treatment temperature the amplitude of the OA decreases monotonically over the whole investigated energy range. To extract the individual contributions to the overall OA spectrum we used a fit procedure including the Urbach exponential tail (which takes into account the intrinsic absorption edge of a-SiO$_2$) and seven Gaussian bands defined by the following peak positions ($P_n$) and FWHMs ($W_n$): $P_1$=4.8 eV and $W_1$=1.05 eV, $P_2$=5.02 eV and $W_2$=0.38 eV, $P_3$=5.80±0.03 eV and $W_3$=0.60±0.05 eV, $P_4$=6.25 eV and $W_4$=0.6 eV, $P_5$=6.8 eV and $W_5$=0.7 eV, $P_6$=7.53±0.01 eV and $W_6$=0.66±0.02 eV, $P_7$=7.98 eV and $W_7$=0.4 eV. The spectral features of the OA bands corresponding to n=1,2,3,6 considered in the fit agree with those attributed in literature to non-bridging oxygen hole center, ODC(II) (unrelaxed O vacancy or divalent Si), E'$_\gamma$ center and ODC(I) (relaxed O vacancy), respectively [1,8]. At variance, the attribution of the OA bands corresponding to n=4,5,7 to specific defects of a-SiO$_2$ is still lacking. For each sample, the exponential profile was obtained by optimizing the fit of the OA spectrum for T=630 K, and then it was fixed in the fit of the spectra relative to the same sample but acquired after thermal treatment above T=630 K. In all the fits, the bands corresponding to n=3 and n=6 were left fully free, whereas the other bands were allowed to change only in amplitude. The errors attributed above to $P_3$, $W_3$, $P_6$ and $W_6$ represent the maximum variability obtained for these parameters by fitting the complete set of OA spectra acquired for our three samples. The results of the fit obtained for the sample QPA thermally treated at T=820 K and that obtained for the sample QC thermally treated at T=630 K are reported in the insets of Fig. 2(a) and 2(b), respectively. As shown, quite a good agreement is found between the fit results and the experimental OA spectra.

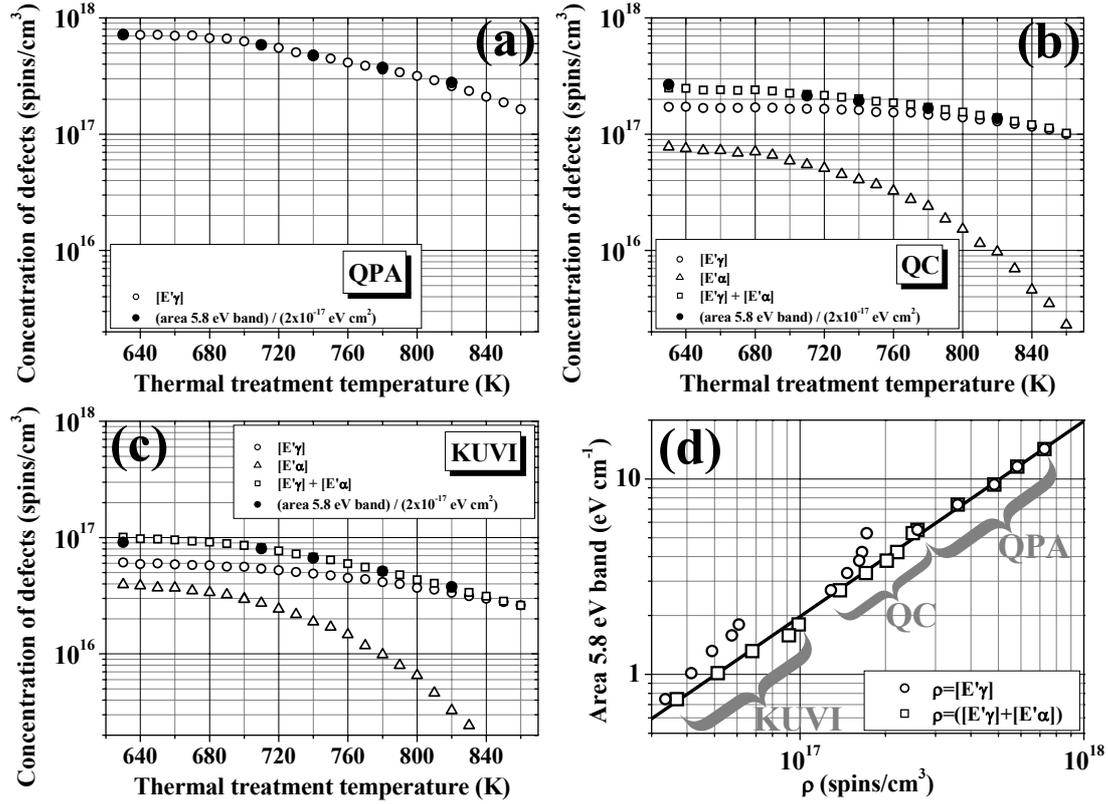

FIG. 3. (a)-(c) Concentration of E'$_\gamma$ and E'$_\alpha$ centers estimated in the samples QPA, QC and KUVI, respectively, compared with the area of the ~5.8 eV OA band divided by $2\times10^{-17}$ eV cm$^2$. In (b)-(c) the sum of concentration ([E'$_\gamma$]+[E'$_\alpha$]) is also reported, for comparison. In (d) the area of the ~5.8 eV OA band measured in the samples QPA, QC and KUVI is compared with the concentration of paramagnetic defects, ρ, for ρ=[E'$_\gamma$] and ρ=([E'$_\gamma$]+[E'$_\alpha$]). The straight line is obtained by a fit to the experimental data, as explained in the text, and is defined by the equation: $y/x = 2\times10^{-17}$ eV cm$^2$. In (a)-(d) the experimental errors are comparable with the symbols dimensions.

In Fig. 3 (a)-(c) the concentrations of E' centers are reported as a function of the isochronal thermal treatment temperature, as estimated in QPA, QC and KUVI, respectively. As shown, in QPA the E'$_\gamma$ centers only are detected, whereas in QC and KUVI both E'$_\gamma$ and E'$_\alpha$ centers are present. Furthermore, it is evident that on increasing the thermal treatment temperature the concentration of both E'$_\gamma$ and E'$_\alpha$ centers gradually anneals out, but with different rates for the two defects. For each sample, the experimental annealing curves of E'$_\gamma$ and E'$_\alpha$ centers were compared with those of the areas of the seven Gaussian bands obtained from the fit of the OA spectra. This study pointed out that only the OA band peaked at ~5.8 eV shows a correlation with the concentration of E' centers in the three samples, whereas the other bands exhibit quite different changes. In fact, in the QC sample the area of the band peaked at 6.25 eV, for example, changes of 15% from T = 710 K to T = 820 K, whereas in the same temperature range the concentration of $E'_\gamma$ changes by 30%, that of $E'_\alpha$ by 80% and that of ($E'_\gamma + E'_\alpha$) by 40%, showing that this band is not correlated to these defects. For this reason the following discussion will be limited to the 5.8 eV OA band.

Since it is well known that the $E'_\gamma$ center is responsible for an OA band peaked approximately at ~5.8 eV [1,6-9], the first step of our analysis was to verify the correlation between the area of the band peaked at ~5.8 eV and the concentration of $E'_\gamma$ centers. These data are shown in Fig. 3(d) (circles). As shown, the area of the ~5.8 eV band is linearly correlated with the concentration of $E'_\gamma$ centers only in the sample QPA, in which no $E'_\alpha$ centers are detected by EPR. At variance, in the samples QC and KUVI, in which both $E'_\gamma$ and $E'_\alpha$ centers are present, the area of the ~5.8 eV band is systematically higher than that expected by extrapolating the data obtained for QPA. This result suggests that a contribution to the area of the ~5.8 eV band could arise also from the $E'_\alpha$ centers. To test this conjecture, the following equation has been considered to fit the experimental data:

$$\text{(area 5.8 eV band)} = \Phi_1 * [E'_\gamma] + \Phi_2 * [E'_\alpha] \quad (1)$$

where the constants $\Phi_1$ and $\Phi_2$ have to be determined. The results of the fit give $\Phi_1=\Phi_2=(2.0\pm0.3)\times10^{-17}$ eV cm$^2$, indicating that Eq. (1) can be simplified as follows

$$\text{(area 5.8 eV band)} = \Phi * ( [E'_\gamma] + [E'_\alpha] ) \quad (2)$$

were $\Phi=\Phi_1=\Phi_2$. The straight line defined by Eq. (2) for $\Phi=2\times10^{-17}$ eV cm$^2$ is compared with the experimental data in Fig. 3(d) (squares). Analogously, in Fig. 3 (b)-(c) the area of the ~5.8 eV band divided by $2\times10^{-17}$ eV cm$^2$ is compared with the sum of concentrations of E' centers ([$E'_\gamma$]+[$E'_\alpha$]). As it is evident from these figures, quite a good agreement is found for all the samples, unequivocally indicating that the $E'_\alpha$ center contributes to the area of the OA band peaked at ~5.8 eV. Since all our attempts to distinguish spectroscopically the bands due to $E'_\gamma$ and $E'_\alpha$ centers were unsuccessful, we attribute to both defects a Gaussian profile with a peak position of 5.80±0.03 eV and a FWHM of 0.60±0.05 eV, as emerged from the fit of the OA spectra. Furthermore, from the estimated values $\Phi_1=\Phi_2=(2.0\pm0.3)\times10^{-17}$ eV cm$^2$ and by using the Lorentz-Lorenz effective field correction [1,8] we obtain the following oscillator strengths: $f_{E'_\gamma} = f_{E'_\alpha} = 0.14\pm0.1$. This value is in excellent agreement with those estimated in previous works for the ~5.8 eV OA band of the $E'_\gamma$ center [1,6,8].

### IV. DISCUSSION

Our results indicate that the $E'_\alpha$ center is responsible for a Gaussian OA band peaked at ~5.80 eV, FWHM of ~0.60 eV and oscillator strength f~0.14. Furthermore, we observe that the OA band of the $E'_\alpha$ center is spectroscopically indistinguishable from that of the $E'_\gamma$ induced in the same samples, supporting the models in which the related electronic transitions involve states highly localized on a structure common to both defects: the O≡Si• moiety [10-12]. On the other hand, our result firmly rules out that the electronic transition could involve processes in which the unpaired electron is transferred from the O≡Si• group to the opposite one, $^+$Si≡O, as proposed by other authors for the $E'_\gamma$ center [13-15]. In this case, in fact, the OA bands associated to $E'_\gamma$ and $E'_\alpha$ centers would be expected to exhibit quite different spectroscopic properties, resulting from the different interatomic distances separating the two Si atoms in these defects [compare Fig. 1(a) and 1(b)], in strong disagreement with our results.

An interesting point which deserves to be discussed concerns a more general comparison of the EPR, optical and structural properties of the $E'_\alpha$ center with those of the $E'_\gamma$ and surface E' center ($E'_s$) [8, 11, 24-26]. The fundamental microscopic structure of the $E'_s$ center is similar to that of the $E'_\gamma$ center but for the fact that in the former the positive $^+$Si≡O group is absent and the unpaired electron projects toward the outside of the material surface [8, 11, 24-26]. The $E'_s$ center is known to possess a $^{29}$Si hyperfine doublet split by ~48 mT [11, 24] and an OA band peaked at ~6.3 eV [11]. By comparing the EPR and OA properties of $E'_\gamma$ and $E'_s$ centers, it is evident that an *increase of both* the hyperfine splitting and the OA band energy peak position occur going from bulk ($E'_\gamma$) to surface ($E'_s$). These differences are generally considered to result from a different O-Si-O angle occurring in the O≡Si• moieties involved in the two defects [25]. In particular, the O-Si-O angle should be larger for the $E'_\gamma$ than for the $E'_s$ [25]. It is worth to note that the EPR and OA properties of the $E'_\alpha$ center does not follow an analogous trend. In fact, our results show that the OA band of the $E'_\alpha$ center is virtually indistinguishable from that of the $E'_\gamma$ center, indicating that a similar O≡Si• moiety with a nearly fixed O-Si-O angle is involved in both defects, whereas their hyperfine splitting differs significantly, being ~49 mT for the $E'_\alpha$ and ~42 mT for the $E'_\gamma$. These properties suggest that the origin of the wider hyperfine splitting of the $E'_\alpha$ center has to be attributed to an higher overall spin localization of the unpaired electron on the Si atom of the O≡Si• group rather than to a smaller O-Si-O angle, as occurs for the $E'_s$ center. In particular, the higher spin localization of the $E'_\alpha$ center could result from both the lacking of the opposite positive $^+$Si≡O group and the repulsive effect exerted by the nearby O atom of the a-SiO$_2$ matrix on the unpaired electron, which limits its partial delocalization [see Fig. 1].

### V. CONCLUSIONS

Here we report a study by OA and EPR on the $E'_\alpha$ point defect in a-SiO$_2$ samples subjected to β-ray irradiation and isochronal thermal treatments. We have found that the $E'_\alpha$ center is responsible for an OA Gaussian band peaked at 5.80±0.03 eV and having a FWHM of 0.60±0.05 eV. The estimated oscillator strength of the related electronic transition is f=0.14±0.1. Our results indicate that this transition, as that attributed to the E'$_\gamma$ center, involves states highly localized on the O≡Si$^\bullet$ moiety.

## AKNOWLEDGEMENTS

Useful discussions with the members of the LAMP group (http://www.fisica.unipa.it//amorphous) are gratefully acknowledged. We thank Dr. B. Boizot of the Ecole Polytechnique (Palaiseau, France) for taking care of the β-ray irradiation in the Van de Graff accelerator. Technical assistance by G. Tricomi and G. Napoli is acknowledged.